\documentclass{elsart}

\usepackage{amssymb}
\usepackage{graphicx}

\begin{document}



\begin{frontmatter}

\title{ \rm Momentum Distributions of Projectile Fragments from Heavy-ion Peripheral Collisions at 15 MeV/nucleon with Emphasis on Trans-Projectile Isotopes}

\author{ S. Koulouris$^{a}$, G.A. Souliotis$^{a,*}$, I. Dimitropoulos$^{a}$, O.Fasoula$^{a}$,}
\author{  K. Palli$^{a}$, M. Veselsky$^{b}$,  A. Bonasera$^{c,d}$ }


\address{ $^{a}$
           Laboratory of Physical Chemistry, Department of Chemistry,
           National and Kapodistrian University of Athens, Athens, Greece }
                 

\address{ $^{b}$ Institute of Experimental and Applied Physics,
                 Czech Technical University, Prague, Czech Republic }

\address{ $^{c}$ Cyclotron Institute, Texas A\&M University,
                     College Station, Texas, USA           }

\address{ $^{d}$ Laboratori Nazionali del Sud, INFN, Catania, Italy }

\address{ $^{*}$ Corresponding author. Email: soulioti@chem.uoa.gr }



\begin{abstract}

This paper presents our recent efforts to study the momentum distributions as well as the production of neutron-rich rare isotopes with heavy-ion beams in the energy region of 15 MeV/nucleon. Experimental cross sections of neutron-rich nuclides from collisions of a $^{86}$Kr (15 MeV/nucleon) beam  with $^{64}$Ni and $^{58}$Ni targets are presented. Experimental data were obtained from the previous work of our group with the MARS mass spectrometer at the Cyclotron Institute of Texas A\&M University. On that note, detailed calculations of yields and momentum distributions of neutron-rich projectile-like fragments are presented for the interaction of $^{86}$Kr with $^{64}$Ni and compared with the aforementioned experimental data.
The calculations were based on a two-step general approach: the dynamical stage of the primary interaction was described with the phenomenological deep-inelastic transfer model (DIT) and the microscopic constrained molecular dynamics model (CoMD); the deexcitation stage of the excited projectile fragments was described with the statistical binary-decay model GEMINI.
The experimental data show an enhancement in the production of neutron-rich isotopes close to the projectile, and interestingly of heavier than the projectile neutron-rich nuclei. The behaviour of the data is relative to the predictions of the CoMD/GEMINI calculation.
The study of the momentum distributions offers a novel route to study the reaction mechanism that dominates the production of the fragments of interest in peripheral heavy-ion collisions at intermediate energies.
%
%
%
In the future, we plan to analyze experimental data that were obtained from the MAGNEX spectometer at the INFN-LNS in Catania, Italy.


\end{abstract}

\end{frontmatter}


\normalsize

\section{Introduction}

During almost a century since the dawn of nuclear physics as an interdisciplinary but also a distinct field in science, approximately one half of the theoretically estimated 7000 bound nuclei have been produced and thoroughly investigated \cite{cite01}. Nuclei far away from the line of beta stability, are not present in nature and have to be prepared in laboratory using appropriate nuclear reactions and separation techniques \cite{cite02}. The exploration of the nuclear landscape toward the neutron drip-line is currently one of the major efforts in nuclear physics research \cite{cite03,cite04,cite05}. The investigation of neutron-rich nuclei offers the possibility to elucidate important astrophysical nucleosynthesis processes, most notably the rapid neutron capture process (r-process), which is responsible for half the abundance of the stable neutron-rich nuclides heavier than iron \cite{cite07,cite08,cite09}.
To this end, the production of very neutron-rich nuclides constitutes 
a central issue in current and upcoming rare isotope beam facilities around the world
(see, e.g., \cite{cite10,cite11,cite12,cite13,cite14,cite15,cite16,cite17,cite18,cite19}).

The traditional routes to produce neutron-rich nuclides are spallation, fission and projectile fragmentation \cite{cite02}.
Spallation is an efficient mechanism to produce rare isotopes for ISOL-type techniques
\cite{cite20}.
Projectile fission is very effective to produce neutron-rich light and
heavy fission fragments (see, e.g., \cite{cite21} for recent efforts on $^{238}$U projectile fission).
Finally, projectile fragmentation offers a universal approach to produce exotic nuclei 
at beam energies above 100 MeV/nucleon (see, e.g., \cite{cite22,cite23}).
This approach is ultimately based on the fact that optimum neutron excess in the fragments is achieved by stripping the maximum possible number of protons
(or a minimum possible number of neutrons).
A novel approach to reach a high neutron excess in the products, apart from the proton striping, is for the projectile nucleus to capture neutrons from the target. Such a possibility is offered by reactions of nucleon exchange  at beam energies
from the Coulomb barrier \cite{cite24,cite25} to the Fermi energy
(20--40 MeV/nucleon) \cite{cite26,cite27}.
Detailed experimental data in this broad energy range are scarce at present \cite{cite25,cite28,cite29}.
In multinucleon transfer and deep-inelastic reactions near the Coulomb barrier \cite{cite25}, the low velocities of the fragments  and the wide angular and ionic charge state distributions may limit the collection efficiency for the most neutron-rich products.
However, the reactions in the Fermi energy regime combine  the advantages of both
low-energy (i.e., near and above the Coulomb barrier) and high-energy (i.e., above 100 MeV/nucleon) reactions. At this energy, the interaction of the projectile with the target enhances the N/Z of the fragments, while the velocities are  high enough to
allow efficient in-flight collection and separation. 
The Fermi energy regime is in fact, a link between dissipative processes observed at low-energy reactions, dominated by mean-field considerations, and high energy collisions for which nucleon-nucleon collisions play an important role \cite{cite30,cite31}.
Our initial experimental studies of projectile fragments from 25 MeV/nucleon reactions  of $^{86}$Kr on $^{64}$Ni \cite{cite26} indicated substantial production of neutron-rich fragments.
Motivated by recent developments in several facilities that will offer either very intense primary beams \cite{cite12,cite15} at this energy range or re-accelerated rare isotope beams \cite{cite10,cite12,cite15,cite16}, we continued our experimental and theoretical studies at 15 MeV/nucleon \cite{cite31,cite32,cite33,cite34}.

In this contribution, after an overview of our experimental measurements with a $^{86}$Kr (15 MeV/nucleon) beam, we present systematic calculations of the production cross sections of projectile fragments with Z = 33 -- 40 and momentum distributions of various products of the reaction $^{86}$Kr (15 MeV/nucleon) with $^{64}$Ni. Concerning the behaviour of the experimental data on the mass distributions and the trend of modern research activity towards the production of heavier than the projectile neutron-rich nuclei in the field of intermediate energies \cite{cite35,cite36,cite37}, special emphasis is given thus to the momentum distributions of those nuclei; the so-called trans-projectile isotopes.
The study of momentum distributions is a novel research activity and is expected to shed light on details of the reaction mechanism. The momentum, as a measure of the energy dissipation of a process, can provide important information on which mechanism ultimately dominates the production of the fragments of interest.
%

%
%

\section{A Brief Presentation of the Theoretical Models}

The calculations performed in this contribution are based on a typical two-stage Monte Carlo approach. The dynamical stage of the collision was described by two models; the phenomenological DIT model and the microscopic CoMD model. In the second stage, the excited projectile-like fragments were deexcited with the binary-decay code GEMINI.
The phenomenological deep-inelastic transfer (DIT) model was implemented by Tassan-Got \cite{cite38} and simulates stochastic nucleon exchange in peripheral and semiperipheral collisions. In DIT model, both the projectile and the target, assumed to be spherical, approach each other along Coulomb trajectories until they are within the range of nuclear interaction. Then, the system is represented as two Fermi gases in contact. The interaction of the projectile and the target leads to formation of a di-nuclear configuration which exists long enough to allow intense exchange through a 'window' in the internuclear potential \cite{cite39}. Transfer of nucleons leads to a gradual dissipation of the kinetic energy of relative motion into internal degrees of freedom (eg. thermal excitation and/or angular momentum). Thus, the interaction of this binary system yields a variation in mass, charge, excitation energy and spin of the primary products. After re-separation, the hot projectile-like and target-like fragments share approximately equal excitation energy and then undergo de-excitation via the GEMINI code. The standard DIT, as will be shown in the results, was not able to properly describe the production cross sections of the neutron-rich fragments in our 15 MeV/nucleon study. As a step toward improvements, we made a proper parametrization of the excitation energy of the primary hot fragments \cite{cite40}. This version of the DIT code, that we call modified DIT (DITmod), will be used in the present contribution, along with the standard DIT model.

The microscopic dynamical model employed in this contribution, is the constrained molecular dynamics (CoMD) model of Bonasera and Papa, which is designed for reactions near and below the Fermi energy \cite{cite41,cite42}. The code is based on the general approach of the quantum molecular dynamics (QMD) models describing the nucleons as localized Gaussian wave packets that interact via an effective nucleon-nucleon interaction. This interaction has a simplified Skyrme-like effective form corresponding to a nuclear matter compressibility of K = 254, for the calculations of this contribution. During the time evolution of the interacting system, a phase space constraint is imposed in order to restore the fermionic nature of the system; thus, effectively restoring the Pauli principle at each time-step of the evolution of the system. In the present CoMD calculations, the dynamical evolution of the system was stopped at t = 600 fm/c (~$2\times 10$\textsuperscript{-21} s). This time interval was optimum, so as to achieve the succesful  completion of the dynamical stage of nucleon transfer, but not that long enough for the deexcitation of hot fragments to take place \cite{cite40}.

To describe the deexcitation of the hot projectile-like fragments produced from the dynamical stage, the binary-decay code GEMINI was implemented. The statistical deexcitation code GEMINI of Charity \cite{cite43,cite44} uses Monte Carlo techniques and the Hauser-Feshbach formalism to calculate the probabilities for fragment emission with Z $<$ 2. Heavier fragment emission probabilities are calculated via the transition state formalism of Moretto. Within this model, the final partition of products is generated by a succession of fragment emissions (binary decays).
\section{Outline of Results and Comparisons} 

A detailed presentation of previously obtained experimental results appear 
in \cite{cite31} in which the mass spectrometric measurements of production 
cross sections of neutron-rich projectile fragments from the reactions of a 15 MeV/ nucleon $^{86}$Kr beam with $^{64,58}$Ni targets are given.
We also point out that the experimental results of the 25 MeV/nucleon reactions 
and the relevant procedures are described in detail in our previous articles \cite{cite26,cite27,cite28,cite29}.    




\begin{figure}[h]                                        
\centering
\includegraphics[width=0.40\textwidth,keepaspectratio=true]{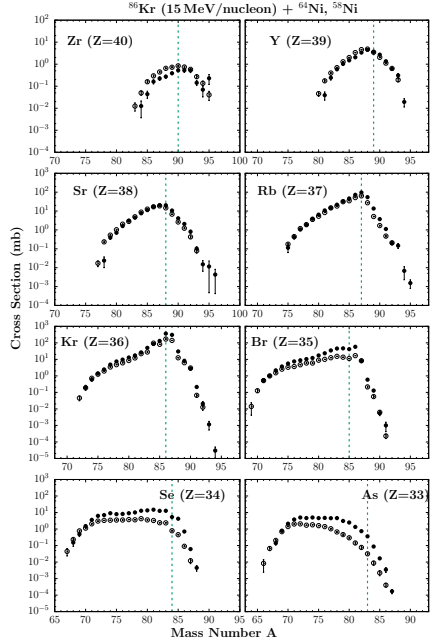}
\caption{ (Color online)
Experimental mass distributions (symbols) of elements Z = 33--40 from the reactions of $^{86}$Kr (15 MeV/nucleon) with $^{64}$Ni and $^{58}$Ni  \cite{cite31}. The data are shown by solid circles for $^{86}$Kr + $^{64}$Ni and open circles for $^{86}$Kr + $^{58}$Ni. The green line indicates the starting point of neutron pickup.
}
\label{fig01}
\end{figure}


In Fig. 1, we present the extracted production cross sections for each isotope of the elements with Z = 33--40 from the reactions $^{86}$Kr (15 MeV/nucleon) with $^{64}$Ni and $^{58}$Ni \cite{cite31}. Solid circles correspond to the first reaction and the open ones to the latter, respectively. The green line indicates the initiation of neutron pickup. As expected, the yields of fragments from the reaction $^{86}$Kr + $^{64}$Ni, are generally larger than the ones of the reaction $^{86}$Kr + $^{58}$Ni due to the relative neutron deficiency of the second target. Firstly we point out that for fragments close to the projectile (e.g., Z = 34, 35) neutron pickup products are observed in both cases with up to six neutrons picked up from the target. As for the Kr isotopes (with Z = 36), we notice the pickup of up to seven neutrons from the target. We observe the production of multinucleon transfer products that picked up several protons and neutrons from the target. Interestingly, in the case of Rb (Z = 37) and Sr (Z = 38), we notice the pickup of up to eight neutrons from the target. 

In Fig. 2, we show the calculated mass distributions of projectile fragments with Z = 33--40 from the reaction $^{86}$Kr (15 MeV/nucleon) with $^{64}$Ni obtained by DIT/GEMINI (solid yellow line), DITmod/GEMINI (solid blue line) and by CoMD/GEMINI (solid green line) and compare them with the experimental data as were described above in Fig.1. In this figure, the experimental data are shown with closed black circles. Starting our discussion with the standard DIT code, we observe that the results of the code at this energy are not satisfactory, especially on the neutron-rich side which is our point of interest. In Fig. 2, the modified DIT (DITmod) code, was able to describe the data adequately on the neutron-rich side, especially for proton-deficient nuclides (eg. isotopes with Z = 32 -- 35). These nuclides come from the projectile nucleus following proton stripping and neutron pickup from the target nucleus. Finally, in Fig. 2, we observe that the results of the CoMD calculations are in overall agreement with the experimental data. A fair agreement with the experimental results is also obtained for the isotopes that are further below the projectile. Interestingly, the calculations of the CoMD model describe quite effectively the neutron-rich side of trans-projectile nuclides. As mentioned above, trans-projectile nuclides have Z higher than the projectile.
These encouraging results, led us to pay special emphasis on the momentum distributions of these trans-projectile isotopes. Due to the trend of modern research activity towards the production of heavier than the projectile neutron-rich nuclei through multinucleon transfer reactions \cite{cite35,cite36,cite37}, special emphasis is given to the present work on their momentum distributions. Understanding the details of the production mechanism of these trans-projectile isotopes is of great importance for the application of such reactions in the production of neutron-rich isotopes of heavy elements \cite{cite45,cite46,cite47}.


\begin{figure}[h]                                        
\centering
\includegraphics[width=0.45\textwidth,keepaspectratio=true]{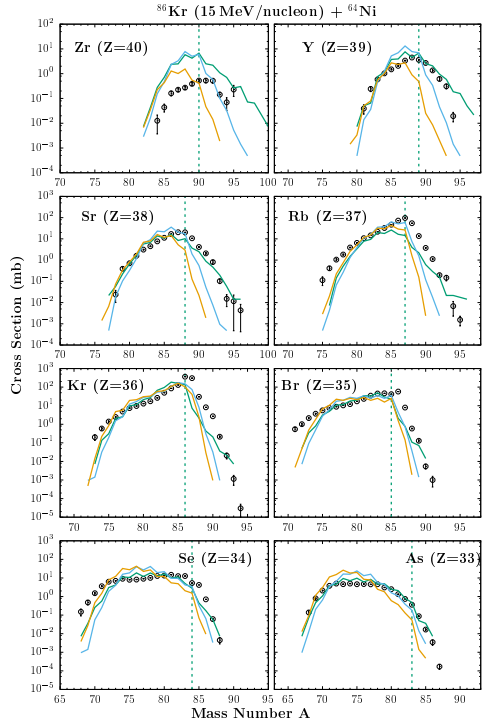} 
\caption{ (Color online) 
   Comparison of calculated mass distributions (lines) of projectile fragments with Z = 33--40 from the reaction of $^{86}$Kr (15 MeV/nucleon) with $^{64}$Ni with the experimental data (closed black circles as in Fig.1) \cite{cite31}. The calculations shown are: DIT/GEMINI (solid yellow line), DITmod/GEMINI (solid blue line) and CoMD/GEMINI (solid green line). The green line indicates the starting point of neutron pickup.}
\label{fig02}
\end{figure}

%
The observable of momentum, is in fact a measure of the energy dissipation caused by the interaction of the projectile-target binary system, and thus can provide important information on which mechanism ultimately dominates the production of the fragments of interest. The general feature of the momentum distributions, as expected, is the presence of two regions: a) a quasielastic peak that corresponds to direct processes, and b) a broad region, located at lower values of P/A (momentum per nucleon) that corresponds to deep inelastic processes and multinucleon transfers. Each frame of the momentum distributions corresponds to a specific isotope produced by the reaction.

\begin{figure}[h]                                        
\centering
\includegraphics[width=0.35\textwidth,keepaspectratio=true]{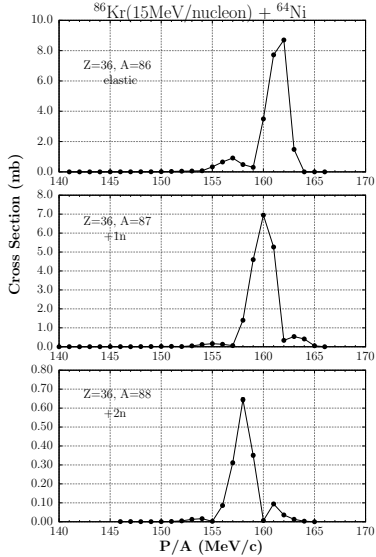} 
\caption{ (Color online) 
   Momentum distributions depicting the elastic channel ($^{86}$Kr) and the products of the pickup of up to two neutrons from the target. The data are shown by solid circles.}
\label{fig03}
\end{figure}

In Fig. 3, the experimental momentum distributions of three isotopes of Kr (Z = 36) from the reaction $^{86}$Kr (15 MeV/nucleon) with $^{64}$Ni are presented. Firstly we notice a quasielastic peak at a P/A value of 163 MeV/c at the elastic channel. However, kinematic calculations carried out by our team, based on the classical two-body kinematics, point out that a  $^{86}$Kr beam of 15 MeV/nucleon should yield P/A = 167.1 MeV/c. So this quasielastic peak is a strong indication of $^{86}$Kr fragments, that are not purely elastic, as they were produced from primary excited fragments, which were finally de-excited via nucleon evaporation, leading to $^{86}$Kr secondary fragments.
Moreover, we observe by the distributions of $^{87}$Kr, $^{88}$Kr (+1n, +2n) that the further pickup of neutrons from the target, leads to considerable dissipation of the produced fragment. In other words, the peaks at the distributions are located at lower values of P/A.



\begin{figure}[h]                                        
\centering
\includegraphics[width=0.35\textwidth,keepaspectratio=true]{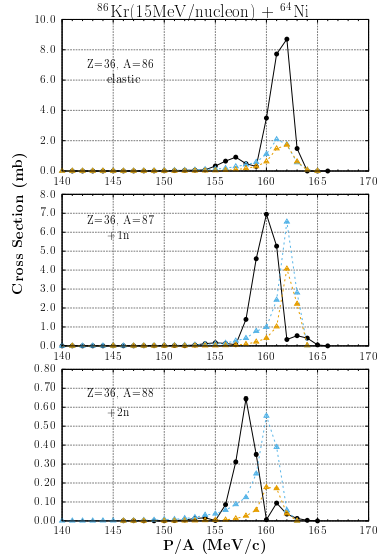} 
\caption{ (Color online) 
   DIT/GEMINI, DITmod/GEMINI calculated momentum distributions depicting the elastic channel ($^{86}$Kr) and the products of the pickup of up to two neutrons from the target. The calculations shown are: DIT/GEMINI (solid yellow line), DITmod/GEMINI (solid blue line). The data are presented by a black line with solid circles.}
\label{fig04}
\end{figure}


\begin{figure}[h]                                        
\centering
\includegraphics[width=0.35\textwidth,keepaspectratio=true]{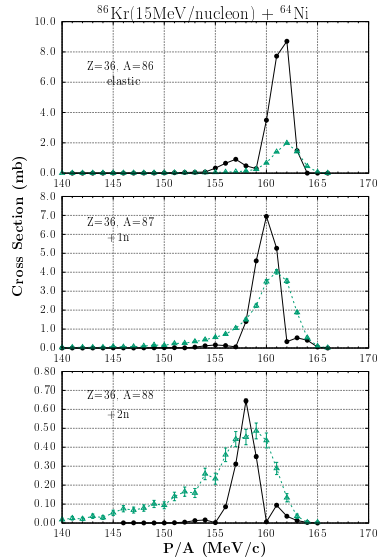} 
\caption{ (Color online) 
   CoMD/GEMINI calculated momentum distributions depicting the elastic channel ($^{86}$Kr) and the products of the pickup of up to two neutrons from the target. The calculations shown are: CoMD/GEMINI (solid green line). The data are presented by a black line with solid circles.}
\label{fig05}
\end{figure}

We now continue with a comparison of the results of the dynamical codes (DIT, DITmod, CoMD) for the same channels as shown in Fig. 3. In Fig. 4, we present the results of both the DIT/GEMINI and DITmod/GEMINI calculations along with the experimental data. The DIT/GEMINI and DITmod/GEMINI calculations are shown by the solid yellow and blue lines respectively, while the experimental data by a black line with solid circles. In general, we can see that the modified DIT (DITmod) code describes the experimental data better than the standard DIT. Nevertheless, fails to describe adequately the elastic channel of this reaction.
In Fig. 5, CoMD/GEMINI calculations along with the experimental data for the same channels are presented. We observe that CoMD describes adequately the distributions of the neutron-pickup products. Although, it cannot describe cannot describe sufficiently the elastic channel of the reaction. Efforts on corrections to the configurations of the code are currently ongoing in order to describe better the data.


\begin{figure}[h]                                        
\centering
\includegraphics[width=0.35\textwidth,keepaspectratio=true]{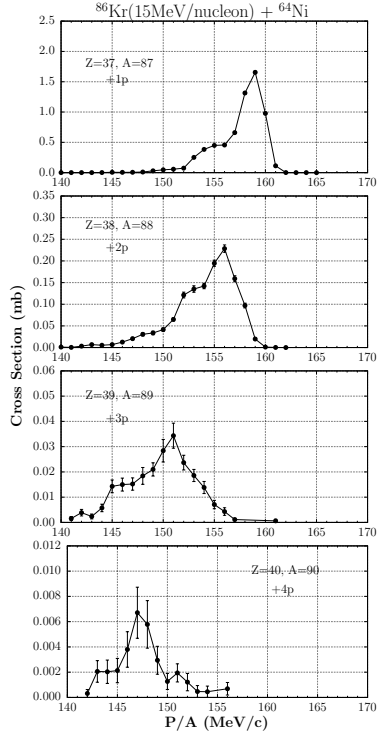}
\caption{ (Color online)
Momentum distributions of the products of the pickup of up to four protons by the projectile. The data are shown by a black line with solid circles.}
\label{fig06}
\end{figure}

In Fig. 6, we present the momentum distributions of $^{87}$Rb, $^{88}$Sr, $^{89}$Y and $^{90}$Zr, which are the products of the capture of up to four protons from the target. We observe that the experimental data of the proton pickup products yield broader distributions in comparison to the neutron pickup products that were examined in Fig. 2. This observation stresses that these final products, are a result of multinucleon transfer from the target. Also, due to the fact that the peaks of these distributions are located in much lower values of P/A, we can point out that the proton pickup leads to a greater dissipation of the system.
Subsequently, we present the results of the DIT, DITmod and CoMD calculations for the pickup of up to four protons from the target. In Fig. 7, the calculations of the standard DIT (yellow line) fail to deliver, in comparison with the experimental data. Nevertheless, the modified DIT (blue line) seems to pave the way towards progress in describing the data. In Fig. 8, the calculations of the CoMD code seem to describe sufficiently the experimental data. Interestingly, in the case of Zr (+4p), we can see a broad neck in the calculations at about 140-145 MeV/c indicates an shows good agreement with the mass distribution of Zr in Fig. 2.

\begin{figure}[h]                                        
\centering
\includegraphics[width=0.25\textwidth,keepaspectratio=true]{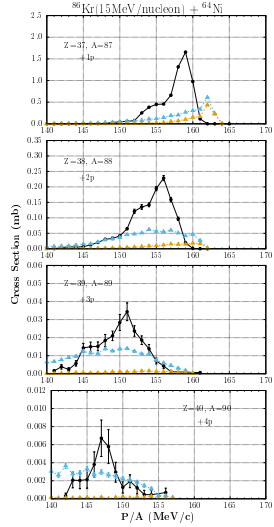}
\caption{ (Color online)
DIT/GEMINI, DITmod/GEMINI calculated momentum distributions depicting the products of the pickup of up to four protons from the target. The calculations shown are: DIT/GEMINI (solid yellow line), DITmod/GEMINI (solid blue line). The data are presented by a black line with solid circles.}
\label{fig07}
\end{figure}

\begin{figure}[h]                                        
\centering
\includegraphics[width=0.30\textwidth,keepaspectratio=true]{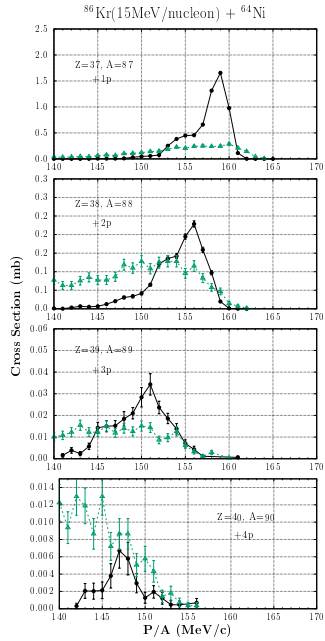}
\caption{ (Color online)
CoMD/GEMINI calculated momentum distributions depicting the products of the pickup of up to four protons from the target. The calculations shown are: CoMD/GEMINI (solid green line). The data are presented by a black line with solid circles.}
\label{fig08}
\end{figure}
\begin{figure}[h]                                        
\centering
\includegraphics[width=0.30\textwidth,keepaspectratio=true]{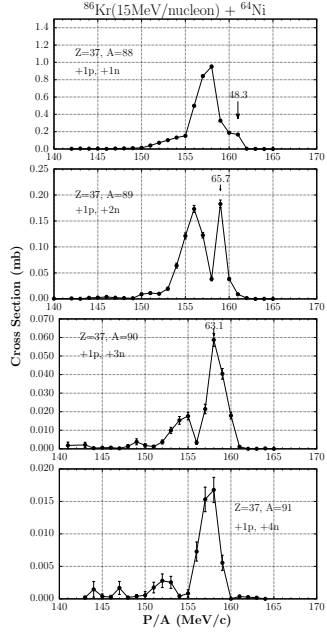} 
\caption{ (Color online) 
Momentum distributions of four neutron-rich isotopes of Rb (Z = 37). The data are shown by a black line with solid circles.}
\label{fig09}
\end{figure}
We now proceed to the study of the momentum distributions of trans-projectile residues. In Fig. 9, we present the experimental momentum distributions of $^{88}$Rb, $^{89}$Rb, $^{90}$Rb and $^{91}$Rb, which are the products of the pickup of a proton and up to four neutrons from the target. Using a simple kinematics code, based on the classical two-body kinematics approach, the total excitation energy of the primary projectile and target fragments for a given value of P/A of the final projectile fragment was calculated and noted on the distributions. We believe that the relatively low values of the total excitation energy of the quasielastic peaks at the first two frames, may be an indication of a direct transfer of $^{2}$H, $^{3}$H clusters. In the next two figures, we present the calculated momentum distributions along with the experimental data for the same channels as presented in Fig. 9. In Fig. 10, we present the results of the DIT, DITmod calculations. It seems that both the standard DIT and the modified DITmod codes, cannot describe the quasielastic peaks of the momentum distributions of the four neutron-rich isotopes of Rb. In general, the results of the CoMD calculations, as shown in Fig. 11, describe better the data, although they tend to an underestimation of the cross sections.
\begin{figure}[h]                                        
\centering
\includegraphics[width=0.25\textwidth,keepaspectratio=true]{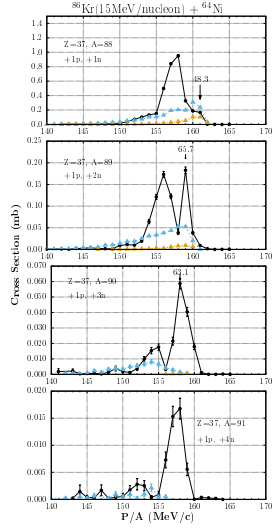} 
\caption{ (Color online) 
DIT/GEMINI, DITmod/GEMINI calculated momentum distributions depicting the four neutron-rich isotopes of Rb (Z = 37). The calculations shown are: DIT/GEMINI (solid yellow line), DITmod/GEMINI (solid blue line). The data are presented by a black line with solid circles.}
\label{fig10}
\end{figure}

\begin{figure}[h]                                        
\centering
\includegraphics[width=0.30\textwidth,keepaspectratio=true]{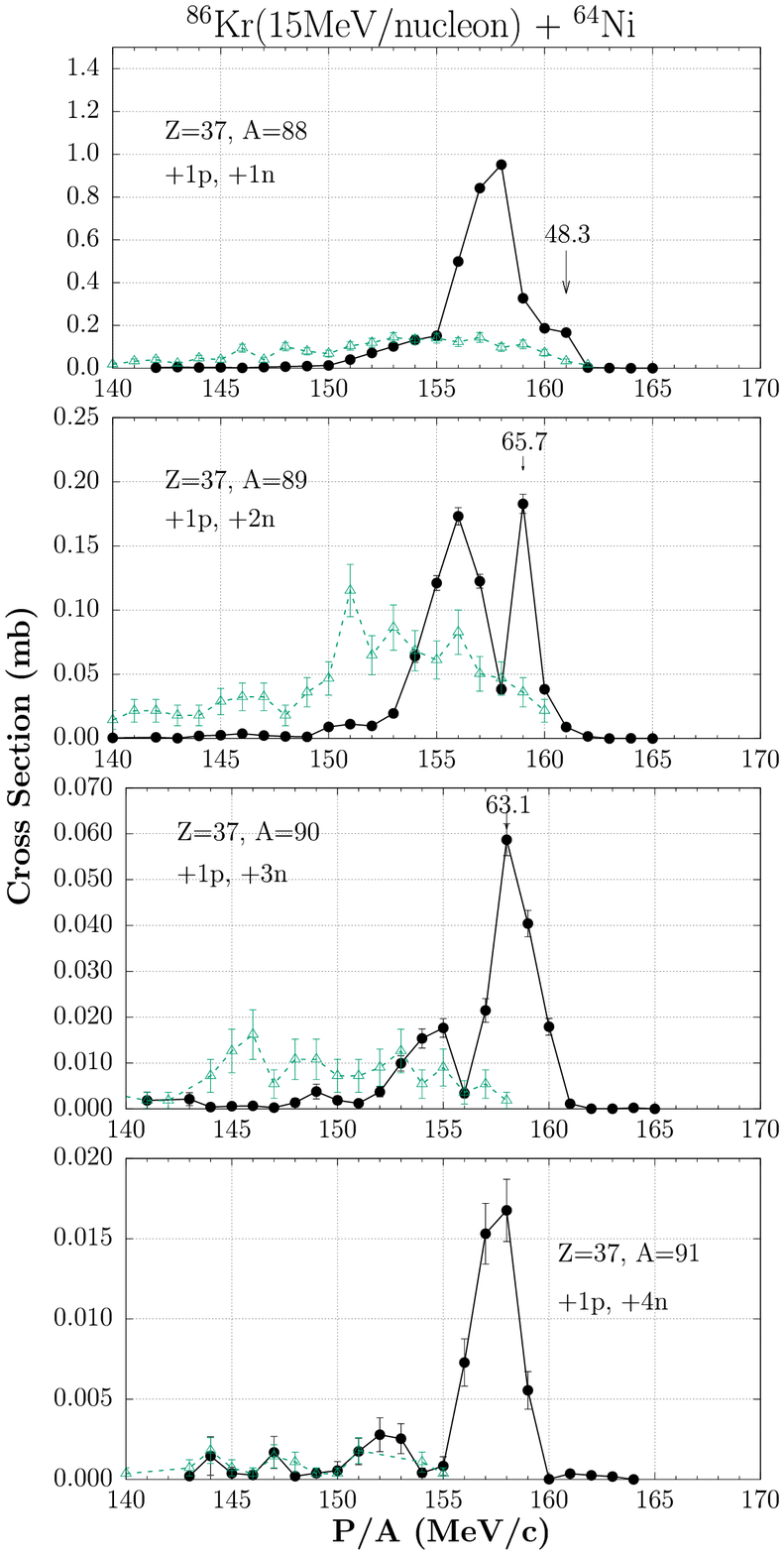} 
\caption{ (Color online) 
CoMD/GEMINI calculated momentum distributions depicting the four neutron-rich isotopes of Rb (Z = 37). The calculations shown are: CoMD/GEMINI (solid green line). The data are presented by a black line with solid circles.}
\label{fig11}
\end{figure}


\begin{figure}[h]                                        
\centering
\includegraphics[width=0.30\textwidth,keepaspectratio=true]{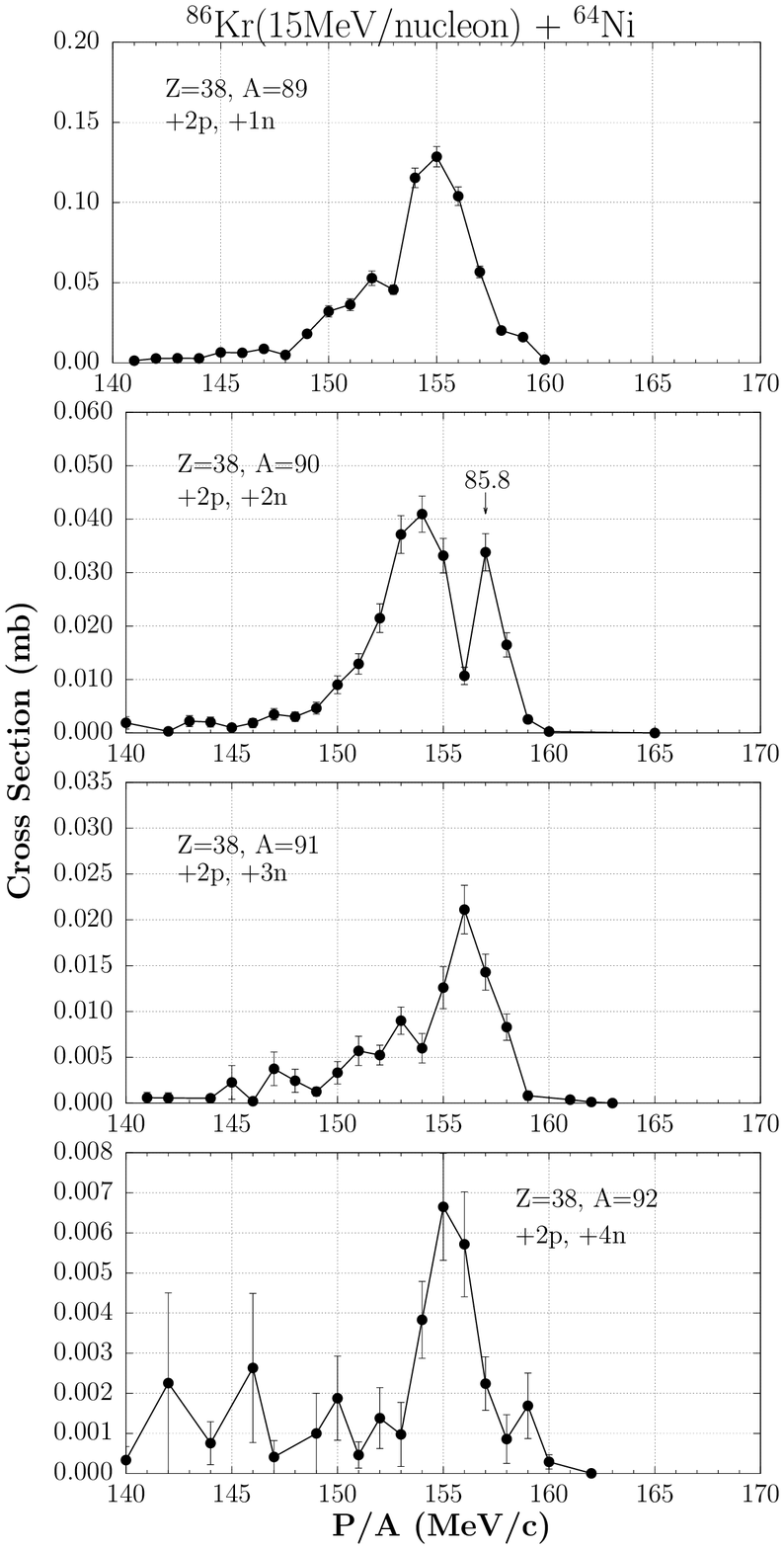} 
\caption{ (Color online) 
Momentum distributions of four neutron-rich isotopes of Sr (Z = 38). The data are shown by a black line with solid circles.}
\label{fig12}
\end{figure}
In Fig. 12, we present the momentum distributions of the four neutron-rich isotopes of Sr (Z = 38). We can see that the peaks of these distributions for each frame are located in lower values of P/A in comparison to the the previous momentum distributions of the Rb isotopes. This shift is a sign of the dissipation of the product due to the further addition of a proton. We also observe that the quasielastic peak of the $^{90}$Sr (+2p, +2n) isotope, located at 157 MeV/c, yields a low total excitation energy of the projectile-like and target-like fragments. This observation indicates a possible direct transfer of an alpha particle from the target to the projectile. We now continue with the results of our dynamical models in Fig. 13 and Fig. 14. From this point on, the standard DIT calculations can't give a description of the experimental data. The modified DITmod seems to describe adequately the dissipative part of the distributions of the data. As far as the CoMD calculations are concerned, it seems that has a satisfactory agreement with the data. Nevertheless, it cannot describe the quasielastic peak of the (+2p, +2n) channel. This is a reasonable observation, as currently the code does not take into account a direct transfer of a cluster nucleus.


\begin{figure}[h]                                        
\centering
\includegraphics[width=0.30\textwidth,keepaspectratio=true]{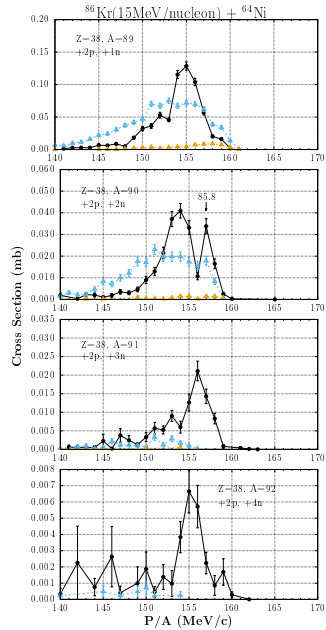} 
\caption{ (Color online) 
DIT/GEMINI, DITmod/GEMINI calculated momentum distributions depicting the four neutron-rich isotopes of Sr (Z = 38). The calculations shown are: DIT/GEMINI (solid yellow line), DITmod/GEMINI (solid blue line). The data are presented by a black line with solid circles.}
\label{fig13}
\end{figure}


\begin{figure}[h]                                        
\centering
\includegraphics[width=0.30\textwidth,keepaspectratio=true]{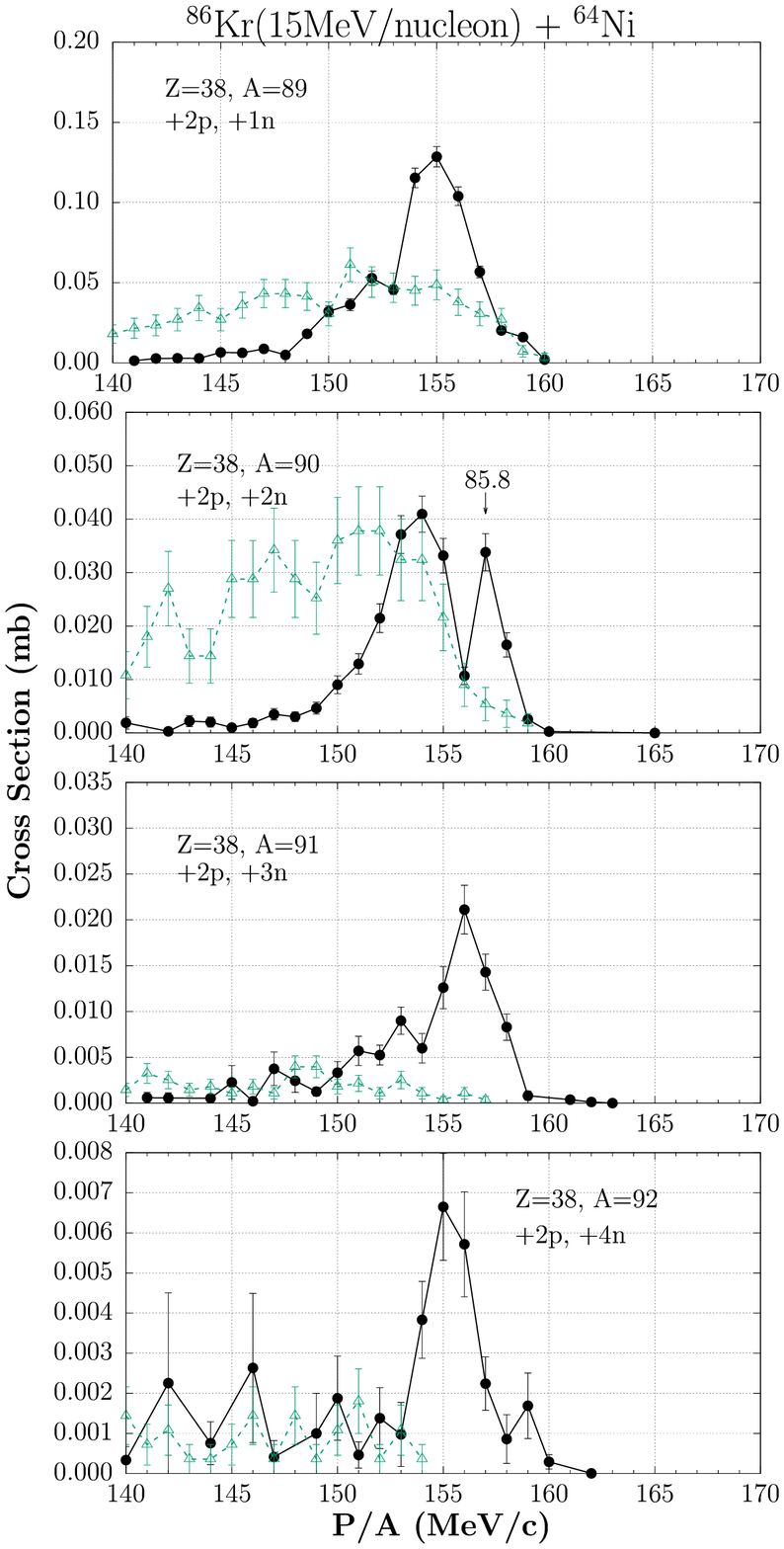} 
\caption{ (Color online) 
CoMD/GEMINI calculated momentum distributions depicting the four neutron-rich isotopes of Rb (Z = 37). The calculations shown are: CoMD/GEMINI (solid green line). The data are presented by a black line with solid circles.}
\label{fig14}
\end{figure}

\begin{figure}[h]                                        
\centering
\includegraphics[width=0.30\textwidth,keepaspectratio=true]{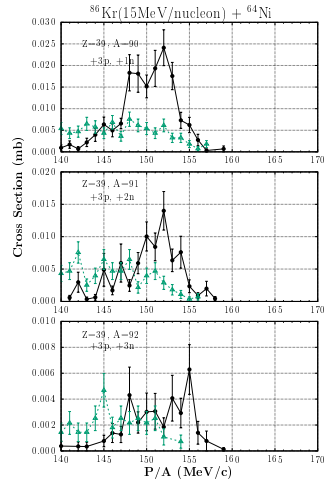} 
\caption{ (Color online) 
CoMD/GEMINI calculated momentum distributions depicting the four neutron-rich isotopes of Y (Z = 39). The calculations shown are: CoMD/GEMINI (solid green line). The data are presented by a black line with solid circles.}
\label{fig15}
\end{figure}
We will now present the final momentum distribution (Fig. 15) for this contribution, which corresponds to the four neutron-rich isotopes of Y (Z = 39). In Fig. 15 we present the experimental data of these channels (black line with solid circles) along with the CoMD/GEMINI calculations (green line). For the isotopes of Y we will not show the results of the DIT/GEMINI and DITmod/GEMINI, as they could not describe at allthe data. We observe that the experimental data of the neutron pickup products of Y yield broader distributions and are located in lower values of P/A in comparison to the previously examined ones. The trend highlights the complexity of the mechanism that led to the production of these exotic nuclides. Finally, we observe that the agreement of the CoMD calculations with the experimental data is reasonable and especially in the (+3p +3n) channel.

\section{Summary and Conclusions}  


We presented recent efforts to study the production of neutron-rich rare isotopes from heavy-ion peripheral collisions at 15 MeV/nucleon and also the experimental production cross sections of neutron-rich nuclides from collisions of a $^{86}$Kr (15 MeV/nucleon)
beam with $^{64}$Ni and $^{58}$Ni targets. As expected, the yields of fragments from the reaction with the neutron-rich target $^{64}$Ni , are generally larger relative to those from the reaction with the neutron-poor $^{58}$Ni.

In parallel, we presented experimental and calculated momentum distributions of specific projectile fragments that were produced by the reaction $^{86}$Kr (15 MeV/nucleon) + $^{64}$Ni, giving special emphasis to the so-called transprojectile residues, as it is well known that multi-nucleon transfer in intermediate energies has currently emerged as one of the best approaches in producing more neutron-rich heavy nuclei. The study of momentum distributions of projectile fragments is done for the first time in the context of our team's overall research on peripheral reactions near the Fermi energy. It is a novel activity as there is limited research worldwide in the kinematic study of reactions in the Fermi energy region through the observable of momentum. The main goal was to shed light on the mechanisms that dominate the production of the reaction products under consideration. It now seems that momentum distributions are a very interesting field of study, which can provide answers to the nature of heavy ion nuclear reactions.
Also we point out that the simultaneous study of the excitation energies of primary fragments could act as a co-diagnostic factor in respect to their energy dissipation, ultimately leading to answers regarding the reaction mechanism.
From the present systematic comparisons, we observed that the macroscopic DIT and DITmod codes cannot describe adequately the experimental data of most channels as seen in the momentum distributions shown, and especially of the transprojectile isotopes. Interestingly, the behaviour of the data is quite relative to the predictions of the CoMD/GEMINI calculation. With proper parametrizations and corrections, we could even confirm the existence of mechanisms such as the direct transfer of an alpha particle from the target to the projectile, as it was suggested in Fig. 12.
In the near future, we plan to analyze experimental data that were obtained from the MAGNEX spectometer at the INFN-LNS in Catania, Italy \cite{cite17}.
We conclude that the study of momentum distributions may lead us to a better understanding of the reaction mechanism at beam energies near the Fermi energy. Thus, a more effective exploitation of nuclear reactions in this regime in order to access extremely neutron-rich isotopes toward the r-process path and the neutron drip-line can be achieved.
Therefore, future experiments in several accelerator facilities can be planned that will enable a variety of nuclear structure and nuclear reaction studies in unexplored regions of the nuclear chart. 


\end{document}